\documentclass[aps,prl,twocolumn,showpacs]{revtex4}
\usepackage{graphicx}
\begin{document}
\draft

\title{ A Universal Phase Diagram for PMN-$x$PT and PZN-$x$PT}


\author{P. M. Gehring}
\address{NIST Center for Neutron Research, National Institute of
  Standards and Technology, Gaithersburg, Maryland 20899}

\author{W. Chen}
\address{Department of Chemistry, Simon Fraser University, Burnaby,
  British Columbia, Canada V5A 1S6}

\author{Z.-G. Ye}
\address{Department of Chemistry, Simon Fraser University, Burnaby,
  British Columbia, Canada V5A 1S6}

\author{G. Shirane}
\address{Physics Department, Brookhaven National Laboratory,
  Upton, New York 11973}

\begin{abstract}
The phase diagram of the Pb(Mg$_{1/3}$Nb$_{2/3}$)O$_3$ and
PbTiO$_3$ solid solution (PMN-$x$PT) indicates a rhombohedral
ground state for $x \le 0.32$.  X-ray powder measurements by Dkhil
{\it et al.} show a rhombohedrally split (222) Bragg peak for
PMN-10\%PT at 80\,K.  Remarkably, neutron data taken on a single
crystal of `the same compound with comparable $q$-resolution
reveal a single resolution-limited (111) peak down to 50\,K, and
thus no rhombohedral distortion.  Our results suggest that the
structure of the outer layer of these relaxors differs from that
of the bulk, which is nearly cubic, as observed in PZN by Xu {\it
et al.}.
\end{abstract}

\pacs{77.84.Dy, 77.80.Bh, 64.70.Kb, 61.12.-q}

\date{\today}
\maketitle

The well-known relaxor Pb(Mg$_{1/3}$Nb$_{2/3}$)O$_3$ (PMN) retains
an average cubic structure down to 5\,K when cooled in zero field
(ZFC).~\cite{Bonneau,deMathan,Ye_review} In this respect PMN
represents a puzzling anomaly among the related complex-perovskite
relaxors PMN-$x$PT and PZN-$x$PT (M=Mg, Z=Zn, PT=PbTiO$_3$), all
of which are believed to exhibit a rhombohedral phase at low
temperatures and low PT
concentrations.~\cite{Choi,Noheda_pmn,Kuwata,Gop_pzn} X-ray
scattering measurements performed on a series of PMN-$x$PT
compounds by Ye {\it et al.}, for example, demonstrate the
presence of a clear rhombohedral distortion for PT concentrations
as low as $x = 0.05$,~\cite{Ye10pt} while Lebon {\it et al}.\ have
reported a detailed x-ray study of the cubic-to-rhombohedral phase
transition in single crystal PZN with high
$q$-resolution.~\cite{Lebon} An interesting comparison between PMN
and PMN-10\%PT was performed by Dkhil {\it et al}.\ using both
x-ray and neutron scattering methods on powder and single crystal
samples which indicate the presence of competing tetragonal and
rhombohedral order.~\cite{Dkhil} In PMN these never result in a
ferroelectric distortion, but in PMN-10\%PT a rhombohedral
distortion is observed below a critical temperature $T_c =
285$\,K.  From these studies, PMN appears to be the exception in
which a ferroelectric phase is never established.

Recent results, however, are now beginning to point towards a
radically different physical picture.  Neutron scattering data
obtained by Ohwada {\it et al}.\ on a single crystal of PZN-8\%PT
suggest the presence of a low-temperature phase in the PZN-$x$PT
family that is not rhombohedral, but that has an average cubic
structure.~\cite{Ohwada}  This new phase was termed ``Phase X.''
Striking evidence of this new phase was subsequently discovered by
Xu {\it et al.} in single crystal PZN where both the rhombohedral
phase and Phase X were observed.~\cite{Xu}  More intriguing is the
fact that the visibility of a given phase depends on the x-ray
energy, and thus the penetration depth.

In this Letter we report the results of a high $q$-resolution
neutron scattering study of single crystal PMN-10\%PT that traces
the evolution of the (111) Bragg peak from the cubic phase above
$T_c = 285$\,K down to 50\,K.  In stark contrast to the findings
of Dkhil {\it et al}.\ we observe no splitting of the (111) peak
at any temperature in PMN-10\%PT, as {\em must} occur under a
rhombohedral transformation.  Further, the (111) peak is nearly
resolution-limited at low temperature, and comparable in width to
that observed by Dkhil {\it et al}.\ at (222) in the cubic phase.
The implications of these results are significant.  Combined with
the results on PZN-8\%PT and PZN obtained by Ohwada {\it et al}.\
and Xu {\it et al}.\ our findings support the diametrically
opposite point of view in which PMN is the rule, not the
exception.  In other words, none of the relaxors in either the
PMN-$x$PT or PZN-$x$PT family transforms to a rhombohedral phase
at low temperatures.  Instead they transform into Phase X.

The neutron scattering data presented here were obtained on the
BT9 triple-axis spectrometer located at the NIST Center for
Neutron Research.  The diffuse scattering near the (300) Bragg
peak was measured at a fixed neutron energy $E_i = E_f =
$14.7\,meV ($\lambda = 2.36$\,\AA) using the (002) reflection of
highly-oriented crystals of pyrolytic graphite (HOPG) as
monochromator and analyzer.  Horizontal beam collimations were
40$'$-46$'$-S-40$'$-80$'$ (S = sample).  Measurements of the (111)
Bragg peak were performed in a non-standard spectrometer
configuration using the (220) reflection of a perfect Ge crystal
as analyzer, tighter beam collimations of
10$'$-46$'$-S-20$'$-40$'$,~\cite{Vinita} and a lower neutron
energy of 8.5\,meV to achieve an extremely sharp $q$-resolution of
0.0018\,rlu (1\,rlu = 2$\pi/$a = 1.5576\,\AA$^{-1}$) FWHM.  The
(220) reflection of Ge was chosen because its $d$-spacing most
closely matches that of the PMN-10\%PT (111) Bragg
reflection.~\cite{Emilio}

A high quality single crystal of PMN-10\%PT was by a top-seeded
solution growth technique.~\cite{Ye}  The growth conditions were
determined from the pseudo-binary phase diagram established for
PMN and PbO.  The crystal, which weighs 2.65\,gm (0.33,cm$^3$), is
an irregular parallelepiped with dimensions $11.3 \times 9.2
\times 4.1 $\,mm$^3$, the largest facets of which are oriented
approximately normal to the cubic [100] direction. At 500\,K the
crystal mosaic measured at (200) is less than 0.06$^{\circ}$
full-width at half-maximum (FWHM).  The crystal was mounted on a
boron nitride post with the [01$\overline{1}$] axis oriented
vertically, giving access to reflections of the form $(hll)$.  The
sample holder assembly was mounted inside the vacuum space of a
high-temperature closed-cycle $^3$He refrigerator that was then
positioned and fixed onto the goniometer of the BT9 spectrometer.

%
%
\begin{figure}
\includegraphics[width=3.0in]{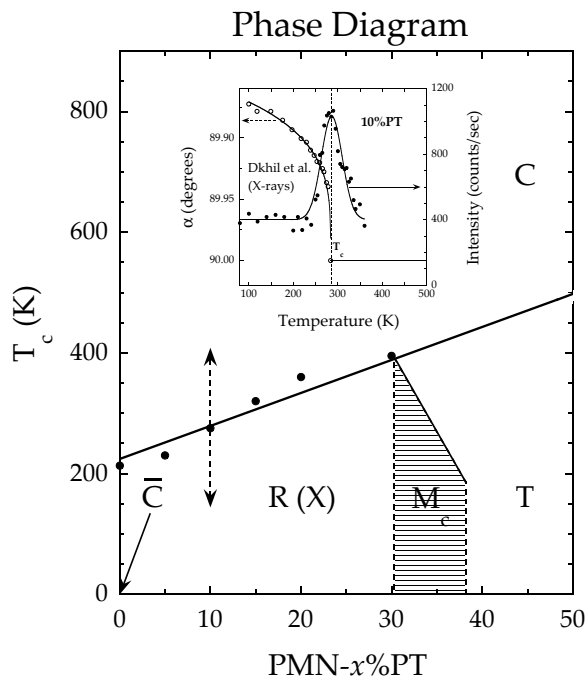}
\caption{\label{fig:1}Currently accepted zero-field phase diagram
of PMN-$x$PT.~\protect\cite{Choi,Noheda_pmn} Inset shows the
temperature dependence of the rhombohedral angle $\alpha$ and
diffuse scattering measured with x-rays by Dkhil {\it et al.} on
PMN-10\%PT (Ref.~\cite{Dkhil}). Lines are guides to the eye.}
\end{figure}
%
%

In Fig.~1 we present the currently accepted phase diagram for
PMN-$x$PT, which shows a cubic (C) to rhombohedral (R) phase
transition for PbTiO$_3$ concentrations $x$ less than $\approx
32$\%.~\cite{Choi,Noheda_pmn}  The solid circles represent
measured values of $T_c$ reported by various authors (see Fig.~4
of Ref.~\cite{Ye10pt}), while the solid line is merely a guide to
the eye.  The lone data point for PMN ($x=0$) corresponds to the
first-order R-to-C phase transition that occurs at 213\,K after
first cooling in a field $E \ge 1.7$\,kV/cm.~\cite{Ye_review}  As
PMN retains an average cubic structure below this temperature when
zero-field cooled (ZFC), we use the notation $\overline{\rm{C}}$
shown by the arrow.  At higher concentrations, PMN-$x$PT exhibits
a narrow region of monoclinic (M$_c$) phase discovered by Noheda
{\it et al.}~\cite{Noheda_pmn} that lies next to the well-known
morphotropic phase boundary (MPB) separating the M$_c$ phase from
the tetragonal (T) phase.  The 10\%PT concentration is indicated
by the vertical dashed line in the phase diagram.

The inset to Fig.~1 shows the temperature dependence of the
rhombohedral angle $\alpha$ (left-hand scale), as well as that of
the diffuse scattering (right-hand scale) near the (300) Bragg
peak ($q$ not specified), for PMN-10\%PT measured by Dkhil {\it et
al} using x-ray diffraction.~\cite{Dkhil} The data for $\alpha$
are consistent with the occurrence of a first-order phase
transition at 285\,K, while the diffuse scattering, which is
characterized by the authors as critical behavior, displays a very
broad peak centered at this same temperature.

%
%
\begin{figure}
\includegraphics[width=3.0in]{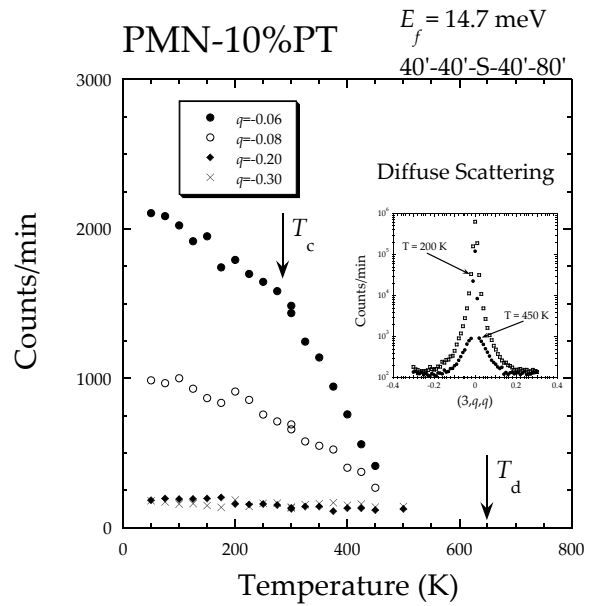}
\caption{\label{fig:2}Neutron diffuse scattering intensity
measured as a function of temperature at $\vec{Q} = (3,q,q)$ on a
PMN-10\%PT single crystal.  The inset shows the full diffuse
scattering profile measured along the [011] direction (transverse
to $\vec{Q}$) at 200\,K and 450\,K.}
\end{figure}
%
%

Neutron diffuse scattering results obtained on the single crystal
of PMN-10\%PT were obtained at the NCNR using relatively coarse
horizontal beam collimations.  These data, shown in Fig.~2 as a
function of temperature, were measured along the [011] direction
near (300) at different values of the reduced wavevector $q$
spanning the range -0.30\,rlu$ \le q \le $0.30\,rlu.  A break in
the slope of the diffuse scattering is evident for $q=-0.06$\,rlu
at a temperature near $T_c$ (shown by the arrow). However there is
no evidence of the peak in intensity at $T_c$ found by x-rays.
Instead, the diffuse scattering intensity increases monotonically
with decreasing temperature down to 50\,K at all values of $q$
studied. These data refute the idea that the diffuse scattering in
PMN-10\%PT is critical in nature. In fact, as shown in the inset
to Fig.~2, the diffuse scattering is quite strong at 450\,K, far
above $T_c = 285$\,K.  The diffuse scattering at this temperature
appears as a very broad peak, with a maximum count rate of just
under 10$^3$ counts/min, superimposed on top of the much narrower
(300) Bragg peak which is roughly two orders of magnitude larger.
In pure PMN the diffuse scattering is known to disappear at the
Burns temperature $T_d = 600 - 650$\,K~\cite{Naberezhnov} above
which polar nanoregions (PNR) no longer exist.~\cite{Burns}  The
approximate corresponding value of $T_d$ for PMN-10\%PT is shown
by the arrow at 650\,K.

Using the high $q$-resolution configuration described earlier,
radial and transverse $q$-profiles of the (111) Bragg peak of the
PMN-10\%PT crystal were measured down to 50\,K.  Two radial scans,
one in the cubic phase at 325\,K, and one in the rhombohedral
phase at 100\,K, are shown side-by-side in Fig.~3.  Above these
two scans are shown the corresponding x-ray scans from Dkhil {\it
et al.}  The horizontal scales for the x-ray data, which were
measured at (222), have been reduced by a factor of two to allow
for direct comparison to the neutron data, which were measured at
(111).  At high temperatures both x-ray and neutron data show the
presence of a single resolution-limited peak, consistent with a
cubic lattice. The horizontal bars represent the calculated BT9
instrumental $q$-resolution FWHM, and show that x-ray and neutron
resolutions are comparable.  At 80\,K, the x-ray data show a clear
splitting of the (222) peak.  At roughly the same temperature,
however, the neutron data show a single peak that is only slightly
broader than resolution.  Remarkably, these data contradict the
finding that PMN-10\%PT is rhombohedral below $T_c$.

%
%
\begin{figure}
\includegraphics[width=3.0in]{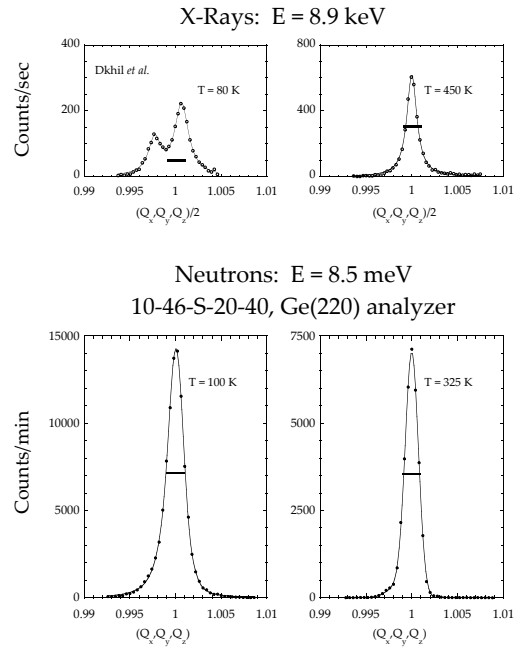}
\caption{\label{fig:3}Comparison of x-ray measurements by Dkhil
{\it et al.} at (222) (top)~\protect\cite{Dkhil} and our neutron
measurements at (111) (bottom) on PMN-10\%PT. The horizontal scale
of the x-ray data have been reduced by a factor of two to allow
direct comparison with the neutron (111) data.  Solid bars
indicate the BT9 instrumental $q$-resolution FWHM.}
\end{figure}
%
%

Our findings strongly support the opposite point of view in which
\emph{none} of the relaxors in the PMN-$x$PT family exhibit a
rhombohedral phase in zero-field below $T_c$. The high-energy
x-ray study of Xu {\it et al}.\ demonstrates the presence of a
$\approx 50$\,$\mu$m thick (1\,$\mu$m = 10$^{-6}$m) outer layer in
the closely related relaxor PZN that undergoes a rhombohedral
distortion at low temperatures, while the interior volume of the
crystal does not.~\cite{Xu}  We believe this to be the case for
PMN-10\%PT as well.  The presence of such an outer layer would
reconcile the x-ray and neutron data shown in Fig.~3 since the
8.9\,keV x-rays penetrate $\approx 10$\,$\mu$m into the sample
whereas neutrons probe the entire crystal volume. It would also
explain why the diffuse scattering peak observed with x-rays at
$T_c$ shown in Fig.~1 is absent when measured with neutrons
(Fig.~2).  Evidence for this outer layer is visible at the base of
the 100\,K neutron data shown in Fig.~3 where an asymmetric
broadening is observed. While these data are insufficient to
extract quantitative information about the thickness of the outer
layer, they can be fitted to two Lorentzian peaks (representing
the rhombohedral phase), and one Gaussian peak (for Phase X).  The
solid line shows a high quality of fit.


In light of the neutron results of Ohwada {\it et al}.\ obtained
on PZN-8\%PT~\cite{Ohwada} and Xu {\it et al}.\ on PZN,~\cite{Xu},
we speculate that there is no bulk zero-field rhombohedral phase
in PZN-$x$PT either.  We propose instead a new and universal phase
diagram for both the PMN-$x$PT and PZN-$x$PT systems, as shown in
Fig.~1, in which the rhombohedral phase is replaced by Phase X.
This scenario is attractive not just because it reconciles the PMN
anomaly, but also because it explains the neutron scattering
results of Wakimoto {\it et al}.\ which demonstrate the presence
of a soft mode, and thus a ferroelectric polarization, below $T_c$
in PMN,~\cite{Wakimoto1} and also of Stock {\it et al}.\ who
showed that PMN and PZN exhibit soft modes with identical
temperature dependences below $T_c$.~\cite{Stock}

While the origin of an outer layer presents an interesting
question, we do not believe it is of fundamental importance to the
underlying properties of the relaxor compounds.  Rather, we feel
that the seminal question concerns the origin of Phase X and its
relationship with the polar nanoregions. Starting from the diffuse
scattering intensity measurements of Vakhrushev {\it et al}.\ on
PMN,~\cite{Vakhrushev} Hirota {\it et al} proposed an entirely new
model of the PNR.~\cite{Hirota}  When ionic displacements due to a
normal ferroelectric distortion are determined from measurements
of Bragg intensities below Tc, the point of reference (origin) for
these displacements is arbitrary. Only the relative positions
between the atoms are meaningful.  But in the case of the diffuse
scattering in PMN, which first appears several hundred degrees
above $T_c$, the origin is not arbitrary because the surrounding
lattice is still cubic.  Following this reasoning, a simple
inspection of the displacements given by Vakhrushev {\it et al}.\
shows that the PNR center-of-mass is shifted from that of
surrounding cubic phase.  An approximate representation of this
shift is shown in Fig.~4.  One logical way to interpret the PMN
displacements is to view them as the sum of a center-of-mass (COM)
conserving component, and a scalar shift $\delta$ that is the same
for all atoms.  Using this approach Hirota {\it et al}.\
demonstrated that the COM-conserving ionic displacements match the
ionic motions associated with the soft TO mode, first observed
above the Burns temperature $T_d$,~\cite{Gehring} thus lending
strong experimental support to this interpretation.  The uniform
shift of the PNR has since been suggested to be the result of the
coupling between the soft TO mode and the TA mode (first
appreciated by Naberezhnov {\it et al}.~\cite{Naberezhnov}).  In
this case the soft coupled TO mode will necessarily carry an
acoustic component that could explain the uniform
shift.~\cite{Yamada,Wakimoto}

The essential point is that the uniform shift of the PNR, which is
of order 60\% that of the Pb displacement, combined with their
large ($\approx 30$ -- 50\AA$^3$) size, makes the process of
\emph{melting} the PNR into the surrounding ferroelectric phase
exceedingly difficult, and thus immensely slow.  In this respect
we believe it is the shift $\delta$ that stabilizes Phase X. While
more study is needed to characterize the PNR, and to determine the
microscopic origin of the displacement $\delta$, we believe that
these ideas form the basis of an elegant and self-consistent model
in which it is possible to understand the basic properties of
these lead-oxide relaxors.


%
%
\begin{figure}
\includegraphics[width=3.0in]{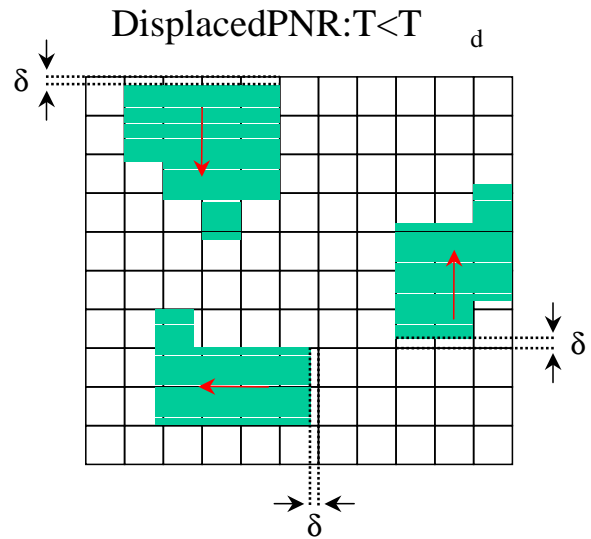}
\caption{\label{fig:4}Schematic diagram of the polar nanoregions
shown in green.  White squares refer to the underlying perovskite
cubic lattice.  The PNR are displaced relative to the cubic
lattice by an amount $\delta$, shown by the dotted lines, in the
direction of the PNR polar axis shown as red arrows.}
\end{figure}
%
%

We thank A.\ Bokov, T.\ Egami, K.\ Hirota, K.\ Ohwada, C.\ Stock,
D.\ Vanderbilt, S.\ Wakimoto, and G.\ Xu, for stimulating
discussions. We also acknowledge financial support from the U.\
S.\ Dept.\ of Energy under contract No.\ DE-AC02-98CH10886, and
the Office of Naval Research, Grant No.\ N00014-99-1-0738.


\begin{references}
\bibitem{Vinita} Our choice of beam collimations were based on Monte Carlo
intensity simulations performed by Vinita Ghosh.
\end{references}
\end{document}